\documentclass[letterpaper, 10 pt, conference]{ieeeconf}  % Comment this line out if you need a4paper

\IEEEoverridecommandlockouts                              % This command is only needed if 
\usepackage{amsmath} % assumes amsmath package installed
\usepackage{amssymb}  % assumes amsmath package installed
\usepackage{amsthm}
\usepackage{tikz}
\usepackage[american]{circuitikz}
\usepackage[linesnumbered,lined,ruled,commentsnumbered]{algorithm2e}

\usepackage{pgfplots}
\pgfplotsset{compat=1.7}
\pgfplotsset{
	/pgfplots/bar  cycle  list/.style={/pgfplots/cycle  list={%
			{orange,fill=orange!30!white,mark=none},%
			{brown!60!black,fill=brown!30!white,mark=none},%
			{lime!40!black,fill=lime,mark=none},%
			{violet,fill=violet!30!white,mark=none},%
			{yellow!30!black,fill=yellow!90!red,mark=none},%
			{red!50!black,fill=red!50!white,mark=none},%
		}
	},
}

\newtheorem{definition}{Definition}

\title{\LARGE \bf
A time series distance measure for efficient clustering of input/output signals by their underlying dynamics*
}

\author{Oliver Lauwers$^{1}$ and Bart De Moor$^{1}$% <-this % stops a space
\thanks{*This work was supported by Belgian Federal Science Policy Office: IUAP P7/19 (DYSCO, Dynamical systems, control and optimization, 2012-2017)
	Flemish Government:
	IWT: TBM IETA(130256); PhD grants
	Industrial Research fund (IOF): IOF Fellowship 13-0260
	VLK Stichting E. van der Schueren: rectal cancer
	EU H2020-SC1-2016-2017 Grant Agreement No.727721: MIDAS Meaningful Integration of Data, Analytics and Services
	KU Leuven Internal Funds C16/15/059, C32/16/013 
	KIC EIT Health: New MOOC - Data Analytics in Health; EIT Health Summer School Innovation on Big Data for Healthy Living
	imec strategic funding 2017. Oliver Lauwers is supported by an SB-grant of the FWO (formerly IWT).}% <-this % stops a space
\thanks{$^{1}$Oliver Lauwers and Bart De Moor. Stadius, Department of Electrical Engineering (ESAT),
        KU Leuven, 3000 Leuven, Belgium
        {\tt\small \{oliver.lauwers,bart.demoor\}@esat.kuleuven.be}}%
}

\begin{document}

\maketitle
\thispagestyle{empty}
\pagestyle{empty}

%%%%%%%%%%%%%%%%%%%%%%%%%%%%%%%%%%%%%%%%%%%%%%%%%%%%%%%%%%%%%%%%%%%%%%%%%%%%%%%%
\begin{abstract}

Starting from a dataset with input/output time series generated by multiple deterministic linear dynamical systems, this paper tackles the problem of automatically clustering these time series. We propose an extension to the so-called Martin cepstral distance, that allows to efficiently cluster these time series, and apply it to simulated electrical circuits data.

Traditionally, two ways of handling the problem are used. The first class of methods employs a distance measure on time series (e.g. Euclidean, Dynamic Time Warping) and a clustering technique (e.g. k-means, k-medoids, hierarchical clustering) to find natural groups in the dataset. It is, however, often not clear whether these distance measures effectively take into account the specific temporal correlations in these time series. The second class of methods uses the input/output data to identify a dynamic system using an identification scheme, and then applies a model norm-based distance (e.g. $H_2$, $H_\infty$) to find out which systems are similar. This, however, can be very time consuming for large amounts of long time series data.

We show that the new distance measure presented in this paper performs as good as when every input/output pair is modelled explicitly, but remains computationally much less complex. The complexity of calculating this distance between two time series of length $N$ is $\mathcal{O}(N\log{N})$.

\end{abstract}

%%%%%%%%%%%%%%%%%%%%%%%%%%%%%%%%%%%%%%%%%%%%%%%%%%%%%%%%%%%%%%%%%%%%%%%%%%%%%%%%
\section{INTRODUCTION}

Time series clustering is an important topic in modern research. State-of-the-art clustering methods of other data types are often not suited for this high-dimensional, temporally correlated data structure. Clustering is the task of finding groups with similar elements in a dataset and consists of three components: a similarity measure based on relevant data features, a clustering algorithm and an evaluation criterion. While the latter two components might carry over, defining a good distance measure is a difficult problem, especially if one is interested in the dynamics of the generating dynamical system of the time series.

Representing the time series as single-input single-output (SISO) linear time invariant (LTI) deterministic dynamical systems further generates problems of its own, as the contributions of the input signal and the impulse response of the system are convolved in the time domain. It is thus not intuitively clear how these two contributions can be separated, for example when one is interested only in the dynamics of the system and not in the specific input signal.

This problem grows ever more relevant as large scale big data time series problems grow more prevalent in areas like finance, medicine, or the industrial internet of things, where clustering is important in tasks like anomaly detection \cite{marti2015anomaly,Vafeiadis}. A typical industrial problem contains several hundred sensors per machine, tens of machines per plant, and several plants per industrial player, collecting data every few seconds, for months or even years of operation time. This results in datasets of several million time points for thousands of series. Clustering techniques should thus scale well.

In Section \ref{sec:existingmethods} we look at state-of-the-art clustering methods for time series from two perspectives, starting from a dataset containing input/output time series pairs, generated by different SISO LTI dynamical systems. From a machine learning point of view, we use an automated clustering method with an off-the-shelf time series distance such as the Euclidean distance or Dynamic Time Warping (DTW). From a system identification point of view, we apply norms such as the $H_2$ or $H_\infty$ norm to compare systems estimated from the data. We find that these techniques either are very fast, but give poor results, or perform well, but are computationally expensive.

Next, in Section \ref{sec:martin}, we look at the Martin cepstral distance \cite{de2002subspace,martin2000metric}, which combines insights from systems theory into a distance measure that can be computed on the raw data. This metric was defined for SISO ARMA models (i.e. LTI models that use white noise as an input signal).

The main contribution of this paper is an extension of the cepstral distance measure, that incorporates deterministic input signals, and allows to calculate distances between a broader class of SISO LTI dynamical systems. It thus allows to cluster time series by dynamics, but remains computationally much simpler than explicitly estimating models.

Subsequently, we apply this new distance measure in Section \ref{sec:application} to an application on electrical circuits, where we generate a dataset consisting of input/output signal pairs, and the problem is to identify which data belong to which generating system. Finally, we conclude the paper and provide some paths for future research in Section \ref{sec:conclusion}.

\section{EXISTING METHODS}
\label{sec:existingmethods}
Existing methods to cluster time series employ a clustering technique, together with some distance measure. The author of \cite{liao2005clustering} discerns three types of distance measures: measures based on raw data, measures based on features of the time series and measures based on models. For the scope of this paper, we will focus on the first and the latter (as the distance measure we propose combines elements of these two broad classes). We present two raw data distance measures, the Euclidean metric and the Dynamic Time Warping metric \cite{keogh2002exact}, and two model-based distance measures, connected to the $H_2$-norm and the $H_\infty$-norm. In the next section, we will introduce and extend the cepstral distance \cite{de2002subspace,martin2000metric}, which combines the efficiency of the raw data distance measures with the insight in generative dynamics of the model norms, and thus has representations both as a raw data distance and as a model-based one.
\subsection{Raw Data Distance Measures}
\label{sub:rawdata}

In what follows we will define $u_m$ to be the input signal of the $m$-th element of a dataset, $y_m$ is the corresponding output signal and $u_m(k)$ or $y_m(k)$ is the value at timepoint $k$ of respectively the input and output of the $m$-th element of the input/output dataset. Time series from element $m$ start at $k = 0$ and end at $k = N_m$. The system that generated an output from a given input will be called the generating (dynamical) system.\\

\subsubsection{Euclidean Distance}

\begin{definition}
	The Euclidean distance, $d_E(\cdot,\cdot)$ treats the time series as a vector, and applies the element-wise Euclidean vector distance between two time series of same length $N_m$, defined as
	\begin{equation}
	d_E(y_m,y_n) = \sqrt{\sum_{k = 0}^{N_m}\left(y_m(k)-y_n(k)\right)^2}.
	\end{equation}
\end{definition}
\begin{description}[\setlabelwidth{$\gamma$}]
\setlength{\itemindent}{-1em}
	\item[\textbf{Advantages}] \hfill
	\begin{itemize}

		\parshape = 2
		1em \dimexpr\columnwidth-1em
		1em \dimexpr\columnwidth-1em
		\item  The Euclidean distance is easy to calculate, allowing for very efficient computation and clustering.
	
		\item No system identification step is needed.
	\end{itemize}
	
	\item[\textbf{Disadvantages}] \hfill 
	
	\begin{itemize}
		\parshape = 2
1em \dimexpr\columnwidth-1em
1em \dimexpr\columnwidth-1em
	\item 	There is no clear link between this distance measure and the generating system.
	
	\item This measure treats the time series as a vector, and ignores the temporal correlations in the data.
	
	\item This measure does not allow to compute distances between time series of different length.
	
	\item This measure does not take the input into account.
	\end{itemize}
	
\end{description}
\
\subsubsection{Dynamic Time Warping}
\ \\

Dynamic Time Warping (DTW) \cite{keogh2002exact,ratanamahatana2004everything} is an algorithm that tries to locally align time series, by \emph{warping} them such that the Euclidean distance between the warped time series is minimal. Mathematically, this \emph{warping}, and the measure that is found in this way, can be described as follows.

Given two output signals, $y_1$ and $y_2$, of length $N_1$ and $N_2$ respectively, a matrix $M$ is constructed, where the $(l,m)$-th element of $M$ is defined as $M_{(l,m)} = (y_1(l) - y_2(m))^2$. A warping path, $W = w_1,w_2,\ldots,w_k,\ldots,w_K$ is then defined, with each $w_k = \left(M_{(l,m)}\right)_k$ an element of matrix $M$ and $max(N_1,N_2) \leq K < N_1 + N_2 - 1$.

The path is subject to the boundary conditions $w_1 = M_{1,1}$ and $w_K = M_{N_1,N_2}$ (i.e. the path starts in one corner of the matrix and ends in the opposite one), has to be continuous, in such a way that two consecutive elements $w_k$ and $w_{k+1}$ are maximally one column and one row apart, and has to be monotonously increasing in its indices, i.e., that in going from $w_k$ to $w_{k+1}$, column nor row number can decrease.

\begin{definition}
We are now interested in the warping path $W_{DTW}$ that minimizes the cost function
\begin{equation}
d_{DTW} \left(y_1,y_2\right) = min\left\{\sqrt{\sum_{k=1}^K w_k}\right\}.
\end{equation}
The sum over this path is then the DTW distance between the time series.
\end{definition}
Though this algorithm is computationally expensive due to the combinatorial nature of the problem, several lower bounds have been devised that can be implemented efficiently. In what follows, we use the Keogh Lower Bound \cite{keogh2002exact} as an efficient approximation to the DTW distance.
\ \\
\begin{description}[\setlabelwidth{$\gamma$}]
\setlength{\itemindent}{-1em}
	\item[\textbf{Advantages}] \hfill 
	
	\begin{itemize}
		\parshape = 2
1em \dimexpr\columnwidth-1em
1em \dimexpr\columnwidth-1em
		\item 	The DTW distance takes into account (part of) the local temporal correlations.
		
		\item No system identification step is needed.
		
		\item Lower bounds on the distance are reasonably efficient.
		
		\item This measure allows to calculate distances between time series of different length.
	\end{itemize}
	
	\item[\textbf{Disadvantages}] \hfill
	
	\begin{itemize}
		\parshape = 2
1em \dimexpr\columnwidth-1em
1em \dimexpr\columnwidth-1em
		\item 	There is no clear link between this distance measure and the generating system.
		
		\item The DTW distance as such is expensive to calculate.
		
		\item This measure does not take the input into account.
	\end{itemize}
	
\end{description}

\subsection{Model-based Distance Measures}

We use the same notation as in subsection \ref{sub:rawdata}. The generating system of the input/output pair $(u_m,y_m)$ will be denoted by $M_m$, and its corresponding transfer function will be written $\mathcal{H}_m$. Based on a model norm $||\cdot||$, the distance between two models $M_i$ and $M_j$ is defined as $||\mathcal{H}_i-\mathcal{H}_j||$.

\subsubsection{$H_2$-norm}

\begin{definition}
	The $H_2$-norm, $||\mathcal{H}||_2$, of a discrete-time system $M$ with transfer function $\mathcal{H}$ is defined as
	\begin{equation}
	||\mathcal{H}||_2 = \sqrt{\frac{1}{2\pi}\int_{-\pi}^{\pi}Tr\left\{\mathcal{H}^H(\textnormal{e}^{i\omega})\mathcal{H}(\textnormal{e}^{i\omega})\right\}d\omega},
	\end{equation}
	where $Tr\{\}$ denotes the trace, the superscript $\cdot^H$ denotes the Hermitian conjugate and $i$ denotes the imaginary unit.
\end{definition}
The $H_2$-norm can be seen as the root-mean-square of the system response to a normalized white noise input. It is thus a measure of the power, or steady-state variance of this response. The $H_2$-norm will be infinite for unstable systems.

\begin{description}[\setlabelwidth{$\gamma$}]
\setlength{\itemindent}{-1em}
	\item[\textbf{Advantages}] \hfill 
	\begin{itemize}
		\parshape = 2
1em \dimexpr\columnwidth-1em
1em \dimexpr\columnwidth-1em
		\item 	The $H_2$-norm provides a physically interpretable way to characterize underlying dynamics of time series.
		
		\item This norm allows to calculate distances between time series of different length.
		
		\item This norm takes the input data into account.
	
	\end{itemize}
	
	\item[\textbf{Disadvantages}] \hfill 
	\begin{itemize}
		\parshape = 2
1em \dimexpr\columnwidth-1em
1em \dimexpr\columnwidth-1em
		\item A system identification procedure is needed, which is both difficult to automate and often computationally expensive (at least more expensive than the raw data measures).
	\end{itemize}
	
\end{description}

\subsubsection{$H_\infty$-norm}

\begin{definition}
	The $H_{\infty}$-norm, $||\mathcal{H}||_\infty$, of a discrete-time system $M$ with transfer function $\mathcal{H}$ is calculated as
	
	\begin{equation}
	||\mathcal{H}||_\infty = \max_{\omega \in [0,\pi[}|\mathcal{H}(\textnormal{e}^{i\omega})|.
	\end{equation}
\end{definition}

This norm thus measures the maximal gain of the frequency response and is called the \emph{gain} of the system. It becomes infinite for systems with poles on the unit circle.
\ \\
\begin{description}[\setlabelwidth{$\gamma$}]
\setlength{\itemindent}{-1em}
	\item[\textbf{Advantages}] \hfill 
	\begin{itemize}
		\parshape = 2
1em \dimexpr\columnwidth-1em
1em \dimexpr\columnwidth-1em
	\item The $H_\infty$-norm provides a physically interpretable way to characterize underlying dynamics of time series.
	
	\item This norm allows to calculate distances between time series of different length.
	
	\item This norm takes the input data into account.
	\end{itemize}
	
	\item[\textbf{Disadvantages}] \hfill 
	\begin{itemize}
		\parshape = 2
1em \dimexpr\columnwidth-1em
1em \dimexpr\columnwidth-1em
		\item A system identification procedure is needed, which is both difficult to automate and often computationally expensive (at least more expensive than the raw data measures).
	\end{itemize}
	
\end{description}

\section{CEPSTRAL DISTANCE}
\label{sec:martin}
In this section we take a closer look at an insightful distance measure on ARMA models, which can be interpreted both as a raw data distance measure and as a model norm: the Martin cepstral norm \cite{de2002subspace,martin2000metric}. We first give a very concise review of the cepstral norm in the stochastic case, then proceed with an extension that allows us to incorporate information about the deterministic input signal.
\subsection{Original Cepstral Norm}
\label{sub:originalmartin}

Based on the power spectral density, $\Phi_y$, of a signal $y$, we can define its power cepstrum, $c_y$ as
\begin{equation}
c_y = \mathcal{F}^{-1}(\log(\Phi_y)),
\label{eq:cepstrumdefinition}
\end{equation}
where $\mathcal{F}^{-1}$ denotes the inverse Fourier transform. This produces a series of coefficients, $c_y(k)$, with integer $k\in[0,N]$, where $N$ denotes the length of time series $y$.

\begin{definition}
	The cepstral norm, $||\mathcal{H}||_C$, of model $M$ with transfer function $\mathcal{H}$, and output $y$ is defined as
	\begin{equation}
	||\mathcal{H}||_C = \sum_{k=0}^{N}k\left(c_{y}(k)\right)^2.
	\label{eq:originalnorm}
	\end{equation}
\end{definition}

For ARMA models it was proven in \cite{de2002subspace} that there are multiple methods to calculate this norm: it can be derived from the subspace angles of the output Hankel matrices of the generating system, from the mutual information of the output space of a system, and from a combination of poles and zeros of the transfer function of the model. Moreover, equation \eqref{eq:originalnorm} allows us to calculate the norm straight from raw data, without the need to identify the underlying systems. We can thus connect the cepstral norm to a raw data distance measure in the following sense:

\begin{definition}
	The cepstral distance, $d_C(y_i,y_j)$, between two time series, $y_i$ and $y_j$, is defined as
	\begin{equation}
	d_C(y_i,y_j) = \sum_{k=0}^{\max\{N_i,N_j\}}k\left(c_{y_i}(k) - c_{y_j}(k)\right)^2,
	\label{eq:original}
	\end{equation}
	where $\max\{N_i,N_j\} - \min\{N_i,N_j\}$ zeros are added at the end of the cepstrum of length $\min\{N_i,N_j\}$.
\end{definition}

%This model interpretation leads to the notation $d_M(y_i,y_j) = d_M(M_i,M_j)$, i.e. the Martin distance between two output time series is the same as the distance between their generating models, and thus constitutes a model norm.
%
%Indeed, in \cite{martin2000metric}, the author proves that for two stable AR models, $M_1$ of order $n_1$, and with poles $\alpha_i \hspace{3pt} (i=1,\cdots,n_1)$ and $M_2$ of order $n_2$, and with poles $\beta_i \hspace{3pt} (i=1,\cdots,n_2)$ it is true that (overbars denoting complex conjugates)
%\begin{equation}
%\begin{aligned}
%d_M(M_1,&M_2) =\\ &\log\frac{\prod_{i=1}^{n_1}\prod_{j=1}^{n_2}|1-\bar{\alpha}_i\beta_j|^2}{\prod_{i,j=1}^{n_1}(1-\bar{\alpha}_i\alpha_j)\prod_{i,j=1}^{n_2}(1-\bar{\beta}_i\beta_j)}.
%\end{aligned}
%\end{equation}

\begin{description}[\setlabelwidth{$\gamma$}]
\setlength{\itemindent}{-1em}
	\item[\textbf{Advantages}] \hfill 
	
	\begin{itemize}
		\parshape = 2
1em \dimexpr\columnwidth-1em
1em \dimexpr\columnwidth-1em
	\item The cepstral distance has an interpretation in terms of the generating model of the time series.
	
	\item The cepstral distance is easy to calculate, allowing for very efficient computation and clustering.
	
	\item No system identification step is needed.
	
	\item This measure allows to calculate distances between time series of different length.
	\end{itemize}

	\item[\textbf{Disadvantages}] \hfill 
	\begin{itemize}
		\parshape = 2
1em \dimexpr\columnwidth-1em
1em \dimexpr\columnwidth-1em
		\item This distance measure can only take information coming from a stochastic input into account.
	\end{itemize}
	
\end{description}

\subsection{Extended Cepstral Distance}
\label{sub:extendedmartin}
The cepstrum, defined in the previous section, finds its roots in homomorphic signal processing \cite[Chapter 10]{oppenheimdigital}. In this type of processing, the original time series data, which often involves complex multiplicative operators like convolutions, is mapped, through a non-linear mapping, to a different domain, that allows for linear filtering. 
The cepstrum, as in equation \eqref{eq:cepstrumdefinition}, is a good example. The convolution in the time domain changes into a multiplication by calculating the power spectral density. Applying a logarithmic transformation then turns the multiplication in frequency domain into an addition. Finally, the inverse Fourier transform takes the problem back to (a transformed version of) the time domain. Equation \eqref{eq:cepstrumdefinition} is thus effectively a method to transform the convolution into an addition.

This allows us to take the output, and separate the contributions from the input signal (which was the main disadvantage left in the cepstral distance, see subsection \ref{sub:originalmartin}) and the impulse responses of the system. Indeed, defining the cepstrum coefficients of the input signal $u$ as $c_u(k)$, and the contribution to the cepstrum coefficients of the transfer function $\mathcal{H}$ as $c_h(k)$, we can write
\begin{equation}
c_y(k) = c_u(k) + c_h(k).
\end{equation}
Based on input/output signal pairs, we now have a measure of the underlying generating system dynamics by looking at $c_h(k) = c_y(k)-c_u(k)$.
\begin{definition}
	The extended cepstral distance, $d_{C_e}((y_i,u_i),(y_j,u_j))$, between two input/output pairs of time series, $(y_i,u_i)$ and $(y_j,u_j)$, with respective transfer functions $\mathcal{H}_i$ and $\mathcal{H}_j$, is defined as
	\begin{equation}
	\begin{aligned}
	d_{C_e}&((y_i,u_i),(y_j,u_j)) = \\
	&\sum_{k=0}^{\min\{N_i,N_j\}}k\left(c_{h_i}(k) - c_{h_j}(k)\right)^2.
	\end{aligned}
	\label{eq:extended}
	\end{equation}
\end{definition}
Note that, for now, this distance measure does not have the whole theoretic framework with connections to subspace angles, mutual information and generating system parameters.\footnote{These theoretical equivalences will be researched and most of them proven to carry over in a forthcoming paper, where we will also try to connect the extended cepstral distance to an extended cepstral model norm.} However, it is clear that the $c_h(k)$ can only come from the generating system dynamics, and thus the distance measure tells us something about these systems, even if it is still unclear what exactly is measured.

We propose this extended cepstral distance as a way to efficiently cluster input/output data by their generating dynamics.
\ \\
\begin{description}[\setlabelwidth{}]
\setlength{\itemindent}{-1em}
	\item[\textbf{Advantages}] \hfill 
	
	\begin{itemize}
		\parshape = 2
1em \dimexpr\columnwidth-1em
1em \dimexpr\columnwidth-1em
	\item The extended cepstral distance is linked to the generating model of the time series.
	
	\item The extended cepstral distance is easy to calculate, allowing for very efficient computation and clustering.
	
	\item No system identification step is needed.
	
	\item This measure allows to calculate distances between time series of different length.
	
	\item This measure takes the input into account.
	\end{itemize}
	
	\item[\textbf{Disadvantages}] \hfill 
	\begin{itemize}
		\parshape = 2
1em \dimexpr\columnwidth-1em
1em \dimexpr\columnwidth-1em
		\item The interpretation of the measure in terms of system parameters and properties is not immediately clear, thus the theoretical framework of the original cepstral distance does not carry over trivially.
	\end{itemize}
	
\end{description}

\section{APPLICATION ON ELECTRICAL CIRCUITS}
\label{sec:application}

\subsection{Simulation Set-Up}

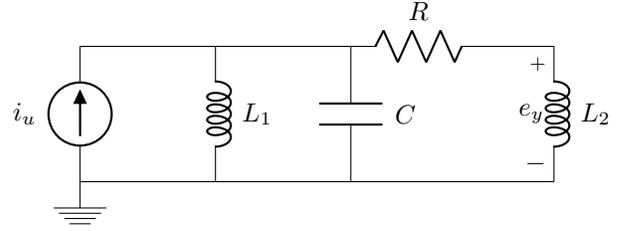
\begin{figure}[t]
	\begin{center}
		\begin{circuitikz}[scale=0.9]
			\draw (0,0)
			node[ground]{}
			to[american current source, i=$i_u$] (0,2) % The voltage source
			to[short] (2,2)
			to[cute inductor=$L_1$] (2,0) % The resistor
			to[short] (0,0);
			\draw (2,2)
			to[short](4,2)
			to[C=$C$] (4,0)
			to[short] (2,0);
			\draw (4,2)
%			to[short] (5,2)
			to[R=$R$] (6,2)
			to[short] (7,2)
			to[cute inductor=$L_2$,v=$e_y$] (7,0)
			to[short] (4,0);
		\end{circuitikz}
		\caption{Electric circuit that was used for the experiments. Two sets, $S_1$ and $S_2$, of values were chosen for the components, namely $S_1 = \{R = 100\Omega,L_1=60\text{H},L_2=20\text{H},C=50\text{F}\}$ and $S_2 = \{R = 100\Omega,L_1=160\text{H},L_2=200\text{H},C=75\text{F}\}$. These two electrical circuits were used to perform the simulations in Section \ref{sec:application}.}
		\label{fig:circuit}
	\end{center}
\end{figure}

To test the proposed techniques, we simulate data coming from electrical circuits. We start out by modelling two circuits with the same topology, but different values for the R, L, and C components. The topology was taken from a course on linear physical systems analysis \cite{LPSA}. The network topology and the values of the components are shown in Figure \ref{fig:circuit}. The input of the system is the current $i_u$, the output is the voltage over $L_2$, $e_y$. State-space models of order 3 are then written down for these networks.

We provide both systems with 200 different input signals (100 outputs of LTI models of order 15, 50 multisine waves corrupted by Gaussian white noise with standard deviation of 0.1 and 50 white noise signals), and measure the output signals. This generates a dataset of 400 input/output signal pairs (200 inputs times 2 models). The question at hand is whether we can use this input/output data, and only this data, to determine which pairs were generated by the same system, i.e. cluster the dataset in two groups, defined by the generative dynamics.

We will do this using the distance measures defined in section \ref{sec:existingmethods} and subsection \ref{sub:originalmartin}, keeping in mind that we use the Keogh Lower Bound \cite{keogh2002exact} as an efficient approximation to DTW. We then compare to the technique developed in subsection \ref{sub:extendedmartin}. There, the power spectral density is estimated by using Welch's method \cite{welch1967use}, which provides a stable approximation\footnote{Note that, for longer time series (i.e. $2^{10}$ and beyond), the Fast Fourier Transform \cite{brigham1988fast} provides a clean enough output to work on. We could thus speed up the algorithm even further for longer series.} of the Fourier transform for short time series. In the Appendix, we give a pseudo-code overview of how the distance measure is calculated, as well as a link to a minimal working example of the simulations discussed and a complexity analysis of the algorithm.

The performance of these simulations will be measured by the Adjusted Rand Index (ARI) \cite{hubert1985comparing,rand1971objective}, which is a similarity measure between partitions. The ARI compares two partitions, $S_1$ and $S_2$, by calculating the ratio of pairs that have the same partitioning status (i.e. belonging to the same partition or not) in both $S_1$ and $S_2$ to the total amount of data pairs, then adjusting the resulting ratio by subtracting the expected value, to account for guessing (i.e. a partitioning that is the result of random guessing is assigned an ARI of 0). An ARI of 1 corresponds to perfectly similar partitions. 

We compare the partitions generated by a hierarchical clustering method, cut-off at two clusters, using distance matrices generated by the different distance measures of section \ref{sec:existingmethods} and section \ref{sec:martin} versus the ground truth (i.e. the time series was generated by the system with parameters $S_1$ or with parameters $S_2$, as in Figure \ref{fig:circuit}).

\subsection{Results}

\begin{figure}[t]
\centering
\hspace{-2.75em}
\begin{tikzpicture}[scale = 0.75]
\pgfplotsset{every axis legend/.append style={
		at={(0.5,1.03)},
		anchor=south}}
\pgfplotsset{
%	/pgfplots/xbar legend/.style={
%		/pgfplots/legend image code/.code={%
%			\draw[##1,/tikz/.cd,bar width=3pt,yshift=-0.2em,bar shift=0pt]
%			plot coordinates {(0cm,0.8em) (2*\pgfplotbarwidth,0.6em)};},
%	},
	/pgfplots/ybar legend/.style={
		/pgfplots/legend image code/.code={%
			\draw[##1,/tikz/.cd,bar width=3pt,yshift=-0.2em,bar shift=0pt]
			plot coordinates {(0cm,0.8em) (2*\pgfplotbarwidth,0.6em)};},
	},
%	/pgfplots/xbar interval legend/.style={%
%		/pgfplots/legend image code/.code={%
%			\draw[##1,/tikz/.cd,yshift=-0.2em,bar interval width=0.7,bar interval shift=0.5]
%			plot coordinates {(0cm,0.8em) (5pt,0.6em) (10pt,0.6em)};},
%	},
%	/pgfplots/ybar interval legend/.style={
%		/pgfplots/legend image code/.code={%
%			\draw[##1,/tikz/.cd,yshift=-0.2em,bar interval width=0.7,bar interval shift=0.5]
%			plot coordinates {(0cm,0.8em) (5pt,0.6em) (10pt,0.6em)};},
%	},
}
\begin{axis}[
%legend pos= outer north east,
%enlargelimits={abs=-0.5},
legend columns = 3,
legend cell align = center,
legend plot pos = left,
ybar=0pt,
ylabel = ARI,
ymin=-0.1,
ymax=1.1,
ymajorgrids,
bar width=0.15,
xlabel = Time Series Length,
xtick={0.5,...,6.5},
xticklabels={$2^6$,$2^8$,$2^{10}$,$2^{12}$,$2^{14}$,$2^{16}$},
x tick label as interval,
xtick pos = left,
ytick pos = left,
%cycle list name = color list,
%fill
]

\addplot+[error bars/.cd,
y dir=both,y explicit]
coordinates {
	(1,-0.00004988) +- (0.0, 0.00000000)
	(2,-0.00004988) +- (0.0, 0.00000000)
	(3,-0.00005185) +- (0.0, 0.00000973)
	(4,-0.00004988) +- (0.0, 0.00000000)
	(5,-0.00004988) +- (0.0, 0.00000000)
	(6,-0.00004988) +- (0.0, 0.00000000)};
\addplot+[error bars/.cd,
y dir=both,y explicit]
coordinates {
	(1,0.00000000) +- (0.0, 0.00000000)
	(2,-0.00000798) +- (0.0, 0.00001838)
	(3,0.00000000) +- (0.0, 0.00000000)
	(4,0.00000000) +- (0.0, 0.00000000)
	(5,0.00000000) +- (0.0, 0.00000000)
	(6,0.00000000) +- (0.0, 0.00000000)};
\addplot+[error bars/.cd,
y dir=both,y explicit]
coordinates {
	(1,-0.00003169) +- (0.0, 0.00005416)
	(2,-0.00004383) +- (0.0, 0.00002936)
	(3,-0.00056914) +- (0.0, 0.00082376)
	(4,-0.00027973) +- (0.0, 0.00060756)
	(5,-0.00020328) +- (0.0, 0.00052319)
	(6,-0.00012421) +- (0.0, 0.00036598)};
\addplot+[error bars/.cd,
y dir=both,y explicit]
coordinates {
	(1,1.00000000) +- (0.0, 0.00000000)
	(2,1.00000000) +- (0.0, 0.00000000)
	(3,1.00000000) +- (0.0, 0.00000000)
	(4,1.00000000) +- (0.0, 0.00000000)
	(5,1.00000000) +- (0.0, 0.00000000)
	(6,1.00000000) +- (0.0, 0.00000000)};
\addplot+[error bars/.cd,
y dir=both,y explicit]
coordinates {
	(1,0.60000201) +- (0.0, 0.49236350)
	(2,0.96000000) +- (0.0, 0.19694639)
	(3,0.68000201) +- (0.0, 0.46882323)
	(4,0.80000000) +- (0.0, 0.40201513)
	(5,0.96000000) +- (0.0, 0.19694639)
	(6,0.96000000) +- (0.0, 0.19694639)};
\addplot+[error bars/.cd,
y dir=both,y explicit]
coordinates {
	(1,0.04000602) +- (0.0, 0.19694515)
	(2,0.40000000) +- (0.0, 0.49236596)
	(3,0.28000000) +- (0.0, 0.45126086)
	(4,0.44000000) +- (0.0, 0.49888765)
	(5,0.76000000) +- (0.0, 0.42923470)
	(6,0.76000000) +- (0.0, 0.42923470)};
\legend{Euclidean,Keogh LB,Cepstral,Extended,$H_2$-norm,$H_\infty$-norm}
%\draw ({rel axis cs:0,0}|-{axis cs:0,0}) -- ({rel axis cs:1,0}|-{axis cs:0,0});
\end{axis}
\end{tikzpicture}
\caption{Performance of the different clustering algorithms, as measured by the ARI. For each time series length, shown on the x-axis, the average ARI over 100 experiments of finding 2 clusters in 400 time series is depicted as the height of the bar. The error bars show the standard deviation for the performance on these 100 experiments. Note that the Euclidean, Keogh LB and cepstral distance have an ARI of 0, i.e., they amount to random guessing. The extended cepstral distance performs best for all series lengths. The model based distances were given a wrong model order, but still give good performance for longer time series.}
\label{fig:ARI}
\end{figure}
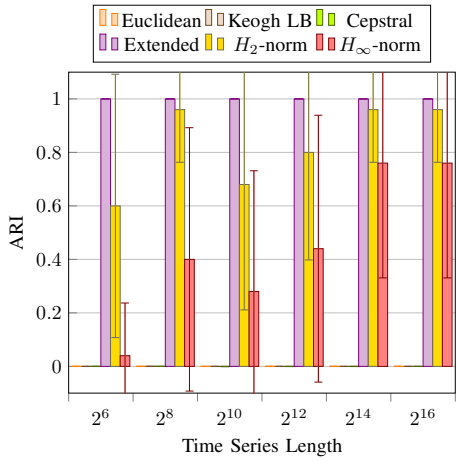

The results for the set-up in the previous subsection are shown in Figure \ref{fig:ARI}, which shows the average and standard deviation for the ARI of the simulation results, and Figure \ref{fig:time}, which shows the average and standard deviation for the execution time of the simulations.

It is clear that the extended cepstral distance gives the best results. In fact, it manages to cluster the simulated input/output pairs perfectly every time. This is, of course, to be expected, as this distance measure was tailored specifically to take into account the dynamics of the underlying model\footnote{We redid the experiments for generating systems of higher order, and the extended cepstral distance still performed best. Results were omitted.}, and nothing but those dynamics. The reasons why it performs better than the other measures will be explained in what follows, and we will again use the distinction between raw data and model-based distances measures from Section \ref{sec:existingmethods}.

\

\subsubsection{Raw Data Distance Measures}
\hfill \\

The reason why the other raw data distance measures do not perform well on the problem at hand, is because they do not take into account the information from the input signal. Indeed, the dynamics of the output are dominated by the input, due to the way the inputs were designed (i.e. the models generating the inputs are of higher order than the models describing the electrical circuits). The other distance measures are thus dominated by contributions coming from the input to cluster the time series, as they cannot separate the different contributions. 

If we only use white noise inputs, we see, on the left hand side in Figure \ref{fig:ARIdistancemeasures}, that the original cepstral distance performs better.\footnote{In fact, the original and extended cepstral distance are equivalent in this case. Indeed, the cepstrum of white noise is only non-zero in its zeroth component, which is not taken into account in the sum in equations \eqref{eq:original} and \eqref{eq:extended}, which coincide in that case.} The Euclidean and DTW distances still do not deliver good results when detecting the difference in dynamics.

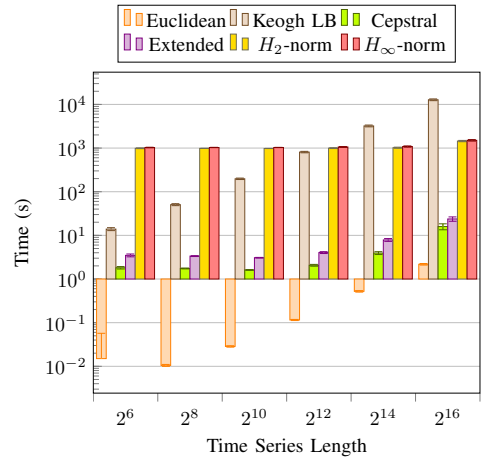
\begin{figure}[t]
	\centering
	\hspace{-2.75em}
	\begin{tikzpicture}[scale = 0.75]
	\pgfplotsset{every axis legend/.append style={
			at={(0.5,1.03)},
			anchor=south}}
	\pgfplotsset{
		%	/pgfplots/xbar legend/.style={
		%		/pgfplots/legend image code/.code={%
		%			\draw[##1,/tikz/.cd,bar width=3pt,yshift=-0.2em,bar shift=0pt]
		%			plot coordinates {(0cm,0.8em) (2*\pgfplotbarwidth,0.6em)};},
		%	},
		/pgfplots/ybar legend/.style={
			/pgfplots/legend image code/.code={%
				\draw[##1,/tikz/.cd,bar width=3pt,yshift=-0.2em,bar shift=0pt]
				plot coordinates {(0cm,0.8em) (2*\pgfplotbarwidth,0.6em)};},
		},
		%	/pgfplots/xbar interval legend/.style={%
		%		/pgfplots/legend image code/.code={%
		%			\draw[##1,/tikz/.cd,yshift=-0.2em,bar interval width=0.7,bar interval shift=0.5]
		%			plot coordinates {(0cm,0.8em) (5pt,0.6em) (10pt,0.6em)};},
		%	},
		%	/pgfplots/ybar interval legend/.style={
		%		/pgfplots/legend image code/.code={%
		%			\draw[##1,/tikz/.cd,yshift=-0.2em,bar interval width=0.7,bar interval shift=0.5]
		%			plot coordinates {(0cm,0.8em) (5pt,0.6em) (10pt,0.6em)};},
		%	},
	}
	\begin{semilogyaxis}[
	%legend pos= outer north east,
	%enlargelimits={abs=-0.5},
	legend columns = 3,
	legend cell align = center,
	legend plot pos = left,
	ybar=0pt,
	ylabel = Time (s),
	ymajorgrids,
	bar width=0.15,
	xlabel = Time Series Length,
	xtick={0.5,...,6.5},
	xticklabels={$2^6$,$2^8$,$2^{10}$,$2^{12}$,$2^{14}$,$2^{16}$},
	x tick label as interval,
	%cycle list name = color list,
	%fill
	xtick pos = left,
	ytick pos = left,
	]
	
	\addplot+[error bars/.cd,
	y dir=both,y explicit]
	coordinates {
		(1,0.01502523) +- (0.0, 0.04205289)
		(2,0.01050697) +- (0.0, 0.00053986)
		(3,0.02856192) +- (0.0, 0.00107219)
		(4,0.11518519) +- (0.0, 0.00341498)
		(5,0.52186870 ) +- (0.0, 0.02062924)
		(6,2.17769534) +- (0.0, 0.08963438)};
	\addplot+[error bars/.cd,
	y dir=both,y explicit]
	coordinates {
		(1,13.93496868) +- (0.0, 1.08063323)
		(2,50.59808369) +- (0.0, 2.62295778)
		(3,197.77850485) +- (0.0, 7.51409654)
		(4,810.14962377) +- (0.0, 30.98372442)
		(5,3216.53100578) +- (0.0, 152.98078828)
		(6,12783.22829556) +- (0.0, 586.30287685)};
	\addplot+[error bars/.cd,
	y dir=both,y explicit]
	coordinates {
		(1, 1.81411497) +- (0.0, 0.11596053)
		(2,1.74571441) +- (0.0, 0.03845423)
		(3,1.61349345) +- (0.0, 0.03132330)
		(4,2.06496827) +- (0.0, 0.09855836)
		(5,3.93670261) +- (0.0, 0.29638088)
		(6,15.91982988) +- (0.0, 2.48249493)};
	\addplot+[error bars/.cd,
	y dir=both,y explicit]
	coordinates {
		(1,3.48495111) +- (0.0, 0.25992700)
		(2,3.35149713) +- (0.0, 0.07586485)
		(3,3.07729191) +- (0.0, 0.05156617)
		(4,4.03768264) +- (0.0, 0.19830723)
		(5,7.86858373) +- (0.0, 0.58257716)
		(6,23.77425429) +- (0.0, 2.71442241)};
	\addplot+[error bars/.cd,
	y dir=both,y explicit]
	coordinates {
		(1,989.78556798) +- (0.0, 18.83113300)
		(2,983.92472604) +- (0.0, 4.22624077)
		(3,979.28355585 ) +- (0.0, 4.73990206)
		(4,996.41536367) +- (0.0, 21.74941411)
		(5,1019.87340339) +- (0.0, 29.07526080)
		(6,1448.21270551) +- (0.0, 49.38834391)};
	\addplot+[error bars/.cd,
	y dir=both,y explicit]
	coordinates {
		(1,1029.81099862) +- (0.0, 14.95718950)
		(2,1030.91302949) +- (0.0, 8.45383926)
		(3,1031.25144539) +- (0.0, 6.53550522)
		(4,1060.09262415) +- (0.0, 22.13229785)
		(5,1084.87930974) +- (0.0, 34.80523710)
		(6,1505.54221839) +- (0.0, 57.92386326)};
	\legend{Euclidean,Keogh LB,Cepstral,Extended,$H_2$-norm,$H_\infty$-norm}
	%\draw ({rel axis cs:0,0}|-{axis cs:0,0}) -- ({rel axis cs:1,0}|-{axis cs:0,0});
	\end{semilogyaxis}
	\end{tikzpicture}
	\caption{Execution time of the different clustering algorithms, measured in seconds. For each time series length, shown on the x-axis, the average time over 100 experiments of finding 2 clusters in 400 time series is depicted as the height of the bar. The error bars show the standard deviation for the execution time on these 100 experiments. Note that the y-axis has logarithmic scale. The extended cepstral distance remains several orders of magnitudes faster than the model-based distances. Note that Keogh LB quickly becomes the computationally most expensive technique. The Euclidean distance is always fastest.}
	\label{fig:time}
\end{figure}

There is thus no hope to achieve better results by taking the input signal into account in the case of the Euclidean distance or the DTW distance. Indeed, the distances look at the shape of the signal, rather than its generative dynamics. DTW is better at this job \cite{keogh2002exact}, but, as we can see from Figure \ref{fig:time}, also has a big disadvantage: it takes a lot of time to compute, especially for long time series, where it even surpasses the model-based distance measures in computation time.

Based on these results, the extended cepstral distance is thus preferred to cluster input/output signals based on the dynamics of their generating models.

\hspace{5pt}

\subsubsection{Model-based Distance Measures}
\hfill \\

\begin{figure}[t]
	\centering
	\hspace{-2.75em}
	\begin{tikzpicture}[scale = 0.75]
	\pgfplotsset{every axis legend/.append style={
			at={(0.5,1.03)},
			anchor=south}}
	\pgfplotsset{
		%	/pgfplots/xbar legend/.style={
		%		/pgfplots/legend image code/.code={%
		%			\draw[##1,/tikz/.cd,bar width=3pt,yshift=-0.2em,bar shift=0pt]
		%			plot coordinates {(0cm,0.8em) (2*\pgfplotbarwidth,0.6em)};},
		%	},
		/pgfplots/ybar legend/.style={
			/pgfplots/legend image code/.code={%
				\draw[##1,/tikz/.cd,bar width=3pt,yshift=-0.2em,bar shift=0pt]
				plot coordinates {(0cm,0.8em) (2*\pgfplotbarwidth,0.6em)};},
		},
		%	/pgfplots/xbar interval legend/.style={%
		%		/pgfplots/legend image code/.code={%
		%			\draw[##1,/tikz/.cd,yshift=-0.2em,bar interval width=0.7,bar interval shift=0.5]
		%			plot coordinates {(0cm,0.8em) (5pt,0.6em) (10pt,0.6em)};},
		%	},
		%	/pgfplots/ybar interval legend/.style={
		%		/pgfplots/legend image code/.code={%
		%			\draw[##1,/tikz/.cd,yshift=-0.2em,bar interval width=0.7,bar interval shift=0.5]
		%			plot coordinates {(0cm,0.8em) (5pt,0.6em) (10pt,0.6em)};},
		%	},
	}
	\begin{axis}[
	%legend pos= outer north east,
	%enlargelimits={abs=-0.5},
	width = 0.4\linewidth,
	height = 0.85\linewidth,
	legend columns = 2,
	legend cell align = center,
	legend plot pos = left,
	ybar=0pt,
	ylabel = ARI,
	ymajorgrids,
	bar width=0.07,
	xtick={0.5,1.5,...,2.5},
	xlabel= Distance Measure,
	x tick label as interval,
	%cycle list name = color list,
	%fill
	xtick pos = left,
	ytick pos = left,
	]
	
	\addplot+[error bars/.cd,
	y dir=both,y explicit]
	coordinates {
		(1,-0.0001) +- (0.0, 0.00001082)};
	\label{Euclidean}
	%	\addlegendentry{Euclidean}
	\addplot+[error bars/.cd,
	y dir=both,y explicit]
	coordinates {
		(1,0.00000000) +- (0.0, 0.00000000)};
	\label{Keogh}
	%	\addlegendentry{Keogh}
	\addplot+[error bars/.cd,
	y dir=both,y explicit]
	coordinates {
		(1,1.0000000) +- (0.0, 0.0000)};
	\label{Original}
	%	\addlegendentry{Original}
	\addplot+[error bars/.cd,
	y dir=both,y explicit]
	coordinates {
		(1,1.00000000) +- (0.0, 0.00000000)};
	\label{Extended}
	%	\addlegendentry{Extended}
	%\legend{Euclidean,Keogh LB,Cepstral,Extended,$H_2$-norm,$H_\infty$-norm}
	%\draw ({rel axis cs:0,0}|-{axis cs:0,0}) -- ({rel axis cs:1,0}|-{axis cs:0,0});
	\end{axis}
	
	\begin{axis}[
	%legend pos= outer north east,
	%enlargelimits={abs=-0.5},
	xshift = 3.5cm,
	width = 0.4\linewidth,
	height = 0.85\linewidth,
	legend columns = 3,
	legend cell align = center,
	%legend plot pos = left,
	legend style={at={(-0.45,1.025)}},
	ybar=0pt,
	ylabel = ARI,
	ymin=-0.1,
	ymax=1.1,
	ymajorgrids,
	bar width=0.07,
	xtick={0.5,1.5,...,2.5},
	xlabel= Model Norms,
	x tick label as interval,
	%cycle list name = color list,
	%fill
	xtick pos = left,
	ytick pos = left,
	]
	
	\addlegendimage{/pgfplots/refstyle=Euclidean}\addlegendentry{Euclidean}
	\addlegendimage{/pgfplots/refstyle=Keogh}\addlegendentry{Keogh}
	\addlegendimage{/pgfplots/refstyle=Original}\addlegendentry{Cepstral}
	
	\pgfplotsset{cycle list shift=3}
	
	\addplot+[error bars/.cd,
	y dir=both,y explicit]
	coordinates {
		(1,1.00000000) +- (0.0, 0.00000000)};
	\addlegendentry{Extended}
	\addplot+[error bars/.cd,
	y dir=both,y explicit]
	coordinates {
		(1,1.00000000) +- (0.0, 0.00000000)};
	\addlegendentry{$H_2$-norm}
	\addplot+[error bars/.cd,
	y dir=both,y explicit]
	coordinates {
		(1,0.9200) +- (0.0, 0.2727)};
	\addlegendentry{$H_\infty$-norm}
	%	\legend{Euclidean,Keogh LB,Cepstral,Extended}
	%\draw ({rel axis cs:0,0}|-{axis cs:0,0}) -- ({rel axis cs:1,0}|-{axis cs:0,0});
	\end{axis}
	\end{tikzpicture}
	\caption{On the left, the performance is shown of the different raw data distance measures, as measured by the Adjusted Rand Index (ARI), in the case of white noise as an input, and time series of length $2^{10}$. Here, the average over 100 experiments with 400 output signals is shown. Note that the original cepstral distance now shows the same performance as the extended one. On the right, results of an experiment where we provided the system identification step with the correct orders of the models are shown. Here, we calculated an average over 100 experiments with 40 output signals, to reduce computation time. Again, we simulated time series of length $2^{10}$. the model-based distances now show better performance.}
	\label{fig:ARIdistancemeasures}
\end{figure}
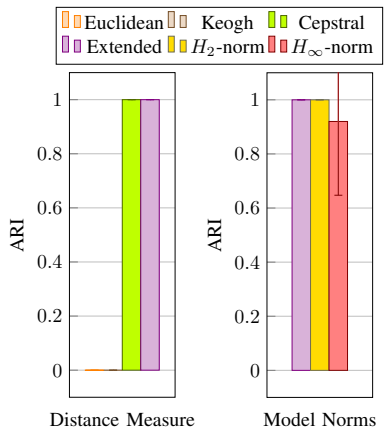

The model-based distance measures show better results than the raw data distance measures, and this again is to be expected. Indeed, the model-based measures take the input information into account and thus manage to peel out the information on the system that generated the input/output pair. However, since a priori we have no information on the order of the underlying system, we arbitrarily have to set a model order. In this case, we estimated transfer functions of order 5. If we share the information on the correct model order (3) with the system identification algorithm, the performance of the model norms increases, as on the right hand side of Figure \ref{fig:ARIdistancemeasures}. 

There exist, of course, schemes to determine appropriate model orders, and more effort can be put in correctly identifying the underlying model. However, as can be seen from Figure \ref{fig:time}, the model norm techniques are already several orders of magnitude slower than the extended cepstrum distance measure. For problems concerning large amounts of long input/output-pairs, as can be found in realistic problems in process industry (see, for example, \cite{marti2015anomaly}, where more than 250 sensors make a measurement every 5 minutes for 6 months), this becomes highly impractical.

The extended cepstral distance is thus preferred over explicitly identifying systems, because of both being easier to automate, and taking less time to compute.

\section{CONCLUSION AND FURTHER RESEARCH}
\label{sec:conclusion}

We have devised a distance measure that is as insightful as a model norm-based distance, yet remains computationally much simpler than explicitly estimating models. It allows to meaningfully cluster large input/output signal pair datasets based exclusively on the dynamics of the generating systems. We have tested it on a simulation of data coming from electrical circuits, where we started from two electrical circuits with a current as input and a voltage difference over an inductor as output. We provided both circuits with 200 different inputs, resulting in 400 input/output pairs.

We then showed that the proposed measure performs as well as model-based distances on estimates of the generative systems, but is much easier to calculate and that other distance measures (Euclidean, DTW) perform much worse.

We furthermore show that, in the stochastic input case, the extended distance proposed in this paper reduces to the original cepstrum distance, which was proven (\cite{de2002subspace,martin2000metric}) to be equivalent to a model norm. This gives hope that the extended distance could also be linked to a model norm. Research that looks into this link is currently under way and will be discussed in a forthcoming paper.

The results indicate the extended cepstral distance measure does a good job of capturing the dynamics of input/output pairs. An application to a real-life dataset is needed to validate the effectiveness in practice, but for the simulated problem at hand, the distance measure succeeded in perfectly distinguishing different dynamics based on raw data alone.

%%%%%%%%%%%%%%%%%%%%%%%%%%%%%%%%%%%%%%%%%%%%%%%%%%%%%%%%%%%%%%%%%%%%%%%%%%%%%%%%

%%%%%%%%%%%%%%%%%%%%%%%%%%%%%%%%%%%%%%%%%%%%%%%%%%%%%%%%%%%%%%%%%%%%%%%%%%%%%%%%

%%%%%%%%%%%%%%%%%%%%%%%%%%%%%%%%%%%%%%%%%%%%%%%%%%%%%%%%%%%%%%%%%%%%%%%%%%%%%%%%
\section*{APPENDIX}
\label{sec:appendix}

A pseudo-code overview of the algorithm is shown in Algorithm \ref{alg:pseudocode}. A minimal working example of the simulations performed in Section \ref{sec:application} is available on GitHub.\footnote{https://github.com/Olauwers/Extended-Cepstral-Distance}

Calculating the extended cepstral distance amounts to estimating the power spectral density of both input and output by Welch's method \cite{welch1967use} (employing the FFT, which is of $\mathcal{O}(n\log{n})$, with $n$ the length of the windows considered in Welch's method), taking the logarithm of the resulting vector, and then applying an inverse Fourier transform (employing the IFFT, running in $\mathcal{O}(N\log{N})$ time, with $N$ the length of the time series) on them. In the end, we then apply a weighted Euclidean distance on the results.

The complexity of calculating the extended cepstral distance between two time series is thus $\mathcal{O}(N\log{N})$, with $N$ the length of the time series.

\begin{algorithm}[t]
	\renewcommand{\algorithmcfname}{\footnotesize Algorithm}
	\footnotesize
	\SetKwInOut{Input}{input}\SetKwInOut{Output}{output}
	
	\Input{Two input/output signal pairs, $\left(y_1,u_1\right)$ of length $N_1$, and $\left(y_2,u_2\right)$ of length $N_2$}
	\Output{The extended cepstral distance $d_{C_e}((y_1,u_1),(y_2,u_2))$ between these two pairs, as defined in Subsection \ref{sub:extendedmartin}}
	%	\BlankLine
	
	\For{$i\leftarrow 1$ \KwTo $2$}{
		$\Phi_{u_i} \xleftarrow[]{\text{Welch's Method}} u_i$\\
		$c_{u_i} \leftarrow \text{ifft}\left(\log\left(\Phi_{u_i}\right)\right)$\\
		%	\BlankLine
		$\Phi_{y_i} \xleftarrow[]{\text{Welch's Method}} y_i$\\
		$c_{y_i} \leftarrow \text{ifft}\left(\log\left(\Phi_{y_i}\right)\right)$\\
		\tcp*[h]{$c_{u_i}$ and $c_{y_i}$ are vectors of length $N_i$}
	}
	%	\BlankLine
	
	$w = \left[0,1,\ldots,\max\{N_1,N_2\}-1\right]$\\
	add $\left(\max\{N_1,N_2\} - \min\{N_1,N_2\}\right)$ 0's to the cepstra of the signal pair of length $\min\{N_1,N_2\}$
	%	\BlankLine
	
	$d_{C_e}((y_1,u_1),(y_2,u_2)) \leftarrow w*\left(\left(c_{y_1} - c_{u_1}\right)^{\intercal} - \left(c_{y_2} - c_{u_2}\right)^{\intercal}\right)^2$
	
	\caption{\footnotesize Algorithm for the extended cepstral distance}
	\label{alg:pseudocode}
\end{algorithm}

%\section*{ACKNOWLEDGMENT}

\bibliography{ClusterBib}

\bibliographystyle{plain}

\end{document}